\documentclass[iopart,longbibliography,superscriptaddress,twocolumn]{revtex4-1}
\usepackage{amssymb}
\usepackage{amsmath}
\usepackage{mathrsfs}
\usepackage{graphicx}
\usepackage{float}
\usepackage[normalem]{ulem}
\usepackage{color}
\usepackage{bm}
\begin{document}
\title{Topological $\pi$ modes and beyond }
\author{Weiwei Zhu}
\email{phyzhuw@gmail.com}
\affiliation{Department of Physics, National University of Singapore, 117551, Singapore}
\author{Jiangbin Gong}
\email{phygj@nus.edu.sg}
\affiliation{Department of Physics, National University of Singapore, 117551, Singapore}
\affiliation{Centre for Quantum Technologies, National University of Singapore, 117543, Singapore}
\author{Raditya Weda Bomantara}
\email{raditya.bomantara@kfupm.edu.sa}
\affiliation{Department of Physics, King Fahd University of Petroleum and Minerals, 31261 Dhahran, Saudi Arabia}
\pagebreak

\maketitle
{Topological states of matter, which have seen rapid development since the last few decades, are characterized by the presence of robust boundary modes. In 2001, Kitaev~\cite{Kitaev2001} identified a particularly important example of such topological modes at the ends of a one-dimensional (1D) $p$-wave superconducting chain (now also known as the Kitaev chain). Due to the presence of particle-hole symmetry in the system, the obtained topological edge modes are pinned at zero energy and are related to the sought-after Majorana particles. Such Majorana zero modes are promising for robust quantum information processing due to their ability to encode qubits nonlocally. Since then, studies of Majorana zero modes and their quantum computing applications have developed into an active research area of their own.}

{In the efforts towards realizing topological matter, Floquet engineering has emerged as a promising approach. It refers to the use of periodic driving to alter an otherwise trivial system into a topologically nontrivial one\cite{Kitagawa2010,Lindner2011}. In 2011, a pioneering study by Jiang \emph{et al.} \cite{Jiang2011} discovered a novel type of Majorana edge modes, termed Majorana $\pi$ modes,  in a periodically driven Kitaev chain.   Unlike the more well-known Majorana zero modes, the driving-induced Majorana $\pi$ modes emerge at half the driving frequency and arise from the interplay among topology, symmetry, and periodic driving. A few years later, several experiments in classical systems, such as photonic \cite{Cheng2019,Wang2022} and acoustic \cite{zhu2022} setups, observed the non-Majorana counterparts of such $\pi$ modes.}

{Motivated by the surge of interest in the studies of such driving-induced boundary modes, this Perspective presents several opportunities that could be achieved in systems supporting topological $\pi$ edge modes. Specifically, we cover their quantum computing applications and their role in the formation of a new phase of matter termed a Floquet time crystal (FTC). Towards the end of this Perspective, we highlight preliminary studies on a generalization of $\pi$ modes, i.e., the $2\pi/k$ modes. Finally, we end this Perspective with brief comments on the future of this topic.}

{\it Definitions and notations.--} In a system with time-periodic Hamiltonian of period $T$, i.e., $H(t+T)=H(t)$, we define the Floquet operator $U$ to be the time-evolution operator over one period. Specifically
\begin{equation}
    U = \mathcal{T} \exp\left(-\frac{i}{\hbar}\int_0^T H(t) dt\right) \;,
\end{equation}
where $\mathcal{T}$ is the time-ordering operator and $\hbar=1$ is assumed hereafter. We further define a Floquet eigenstate $|\varepsilon\rangle$ and its quasienergy $\varepsilon\in \left(-\pi/T,\pi/T\right]$ from the eigenvalue equation $U|\varepsilon\rangle = e^{-i\varepsilon T} |\varepsilon\rangle$. Let us also define a quasienergy $\epsilon$ excitation $\gamma_\epsilon$ to satisfy $U\gamma_\epsilon U^\dagger = e^{-i \epsilon T} \gamma_\epsilon $. Physically, $\gamma_\epsilon$ maps $|\varepsilon\rangle$ to another Floquet eigenstate $|\varepsilon+\epsilon\rangle \propto \gamma_\epsilon |\varepsilon\rangle$. In particular, Majorana zero modes and Majorana $\pi$ modes are respectively defined as quasienergy $0$ and $\pi/T$ excitations, which further satisfy the Hermiticity condition $\gamma_{0/\pi}^\dagger =\gamma_{0/\pi}$.

{{\it Quantum computing applications.--} The discovery of Majorana $\pi$ modes in systems which could also host Majorana zero modes raised an important question about their implications in topological quantum computation. This aspect was first explored by Bomantara and Gong \cite{Raditya2018} via a scheme for simulating the process of braiding between a Majorana zero mode and a Majorana $\pi$ mode in a strictly 1D setup. Such a braiding process, which is necessary for operating various quantum gates on the nonlocal qubits encoded by Majorana modes, typically requires a movement in two physical dimensions, or at least in a quasi-one dimension~\cite{Alicea2011}. In \cite{Raditya2018}, accomplishing this process in an even lower dimension is made possible by exploiting the time dimension and the quasienergy difference between Majorana zero modes and Majorana $\pi$ modes.}

\begin{figure*}
\includegraphics[width=0.95\linewidth]{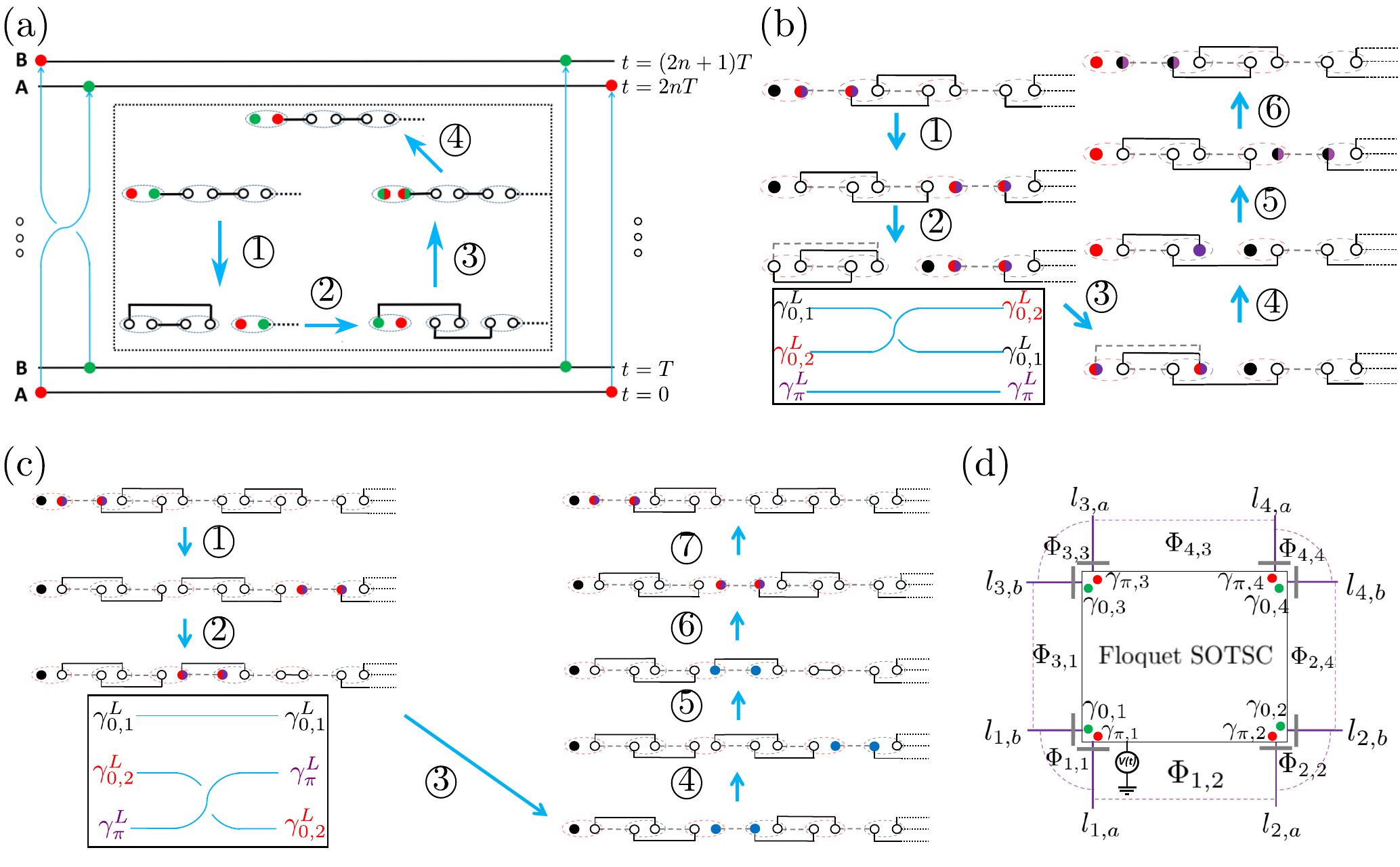}
\caption{(a) Schematics of a protocol for braiding a Majorana zero mode and a Majorana $\pi$ mode in \cite{Raditya2018}.
(b) Implementation of the phase gate via braiding of Majorana modes in the system of \cite{Raditya2018a}.
(c) Implementation of the Hadamard gate via braiding of Majorana modes in the system of \cite{Raditya2018a}.
(d) A minimal setup for measurement-based quantum computations with Majorana zero modes and Majorana $\pi$ modes in a periodically driven second-order topological superconductor \cite{Raditya2020}.
}
\label{pimode}
\end{figure*}

{To elaborate the above mechanism, consider a periodically driven 1D Kitaev chain in which two Majorana operators $\gamma_j^A$ and $\gamma_j^B$ occupy the $j$th site (roughly speaking, $\gamma_j^A$ and $\gamma_j^B$ represent the real and imaginary parts of the fermionic creation operator at site $j$, respectively).   By referring to \cite{Raditya2018} for the full Hamiltonian, it is found that one of its Majorana zero modes and Majorana $\pi$ modes can be explicitly written as a symmetric and anti-symmetric superposition of $\gamma_1^A$ and $\gamma_1^B$. In this case, the task of braiding the two Majorana modes can be equivalently formulated in terms of braiding $\gamma_1^A$ and $\gamma_1^B$. To achieve the latter, adiabatic deformation of the Hamiltonian is systematically devised to map $\gamma_1^A\rightarrow \gamma_1^B$ and $\gamma_1^B\rightarrow -\gamma_1^A$, as schematically depicted in Fig.~\ref{pimode}(a).}

{Note that while $\gamma_1^A$ and $\gamma_1^B$ are already exchanged at the end of step 2 in Fig.~\ref{pimode}(a), the system Hamiltonian has not returned to its original form. To rectify this, an additional step 3 is executed, which amounts to adiabatically deforming the system parameters back to their original values. In doing so, however, the key idea of \cite{Raditya2018} is to only vary the system parameters every other period. This in turn generates a non-Abelian geometric phase on the subspace spanned by the Majorana zero mode and Majorana $\pi$ mode, which prevents $\gamma_1^A$ and $\gamma_1^B$ to be exchanged back during this additional step. In particular, $\gamma_1^A$ and $\gamma_1^B$ are mapped into their symmetric and anti-symmetric superpositions at the end of step 3, thus corresponding to half braiding. To achieve the full braiding between $\gamma_1^A$ and $\gamma_1^B$, the same steps 1-3 are then repeated one more time (see Fig.~\ref{pimode}(a)).}

{Apart from reducing the physical resources required to execute braiding, the half-braiding protocol of \cite{Raditya2018} is particularly important for generating a magic state~\cite{Bravyi2005} that enables a universal quantum computation. With Majorana zero modes alone, a magic state is typically generated by a dynamical process~\cite{Freedman2006,Karzig2016} that requires a precise control over system parameters. By contrast, by including Majorana $\pi$ modes into the picture, a geometrically protected magic state generation could be achieved.}

{In a follow up work by Bomantara and Gong \cite{Raditya2018a}, the same approach was applied to another periodically driven 1D Hamiltonian that supports two Majorana zero modes and one Majorana $\pi$ mode at each end. Taking into account both ends of the system, the presence of six Majorana modes in total allows for the encoding of two qubits under the constraint of fermion parity conservation. Braiding different pairs of Majorana modes (see, e.g., Fig.~\ref{pimode}(b,c)) then leads to a variety of quantum gate operations acting on these two qubits, as comprehensively presented in \cite{Raditya2018a}.}

{Another braiding protocol in a strictly 1D periodically driven Kitaev chain was developed by Bauer \emph{et al.}~\cite{Bauer2019}. Similar to the protocol of \cite{Raditya2018}, the presence of Majorana $\pi$ modes and additional $2T$-periodic parameters are key elements in its construction. However, rather than utilizing the $2T$-periodic parameters for the adiabatic deformation as was done in \cite{Raditya2018}, the former are introduced directly as extra terms in the Hamiltonian near the center of the chain. This allows for a Majorana zero mode and a Majorana $\pi$ mode to hybridize at the center of the chain, which in turn opens a path for another Majorana mode to be moved from one end to the other. By further moving either the Majorana zero mode or the Majorana $\pi$ mode from the center to the other end of the chain, the braiding process is completed.}

{In 2020, further progress was made in \cite{Raditya2020}, which develops a means to braid different pairs of Majorana zero modes and/or Majorana $\pi$ modes without physically moving them. This is accomplished through a series of parity measurements between pairs of Majorana modes. Such measurements are each obtained by a parity dependent conductance that follows a Mach-Zehnder interference-like mechanism \cite{Raditya2020}. Recently, Matthies \emph{et al.}~\cite{Matthies2022} propose a means to prepare a Floquet eigenstate in the system of \cite{Raditya2020}, thus bringing it one step closer towards a potential experimental realization.}

We also highlight the work by Tong \emph{et al.}~\cite{Tong2013} that demonstrated the possibility for multiple Majorana zero modes and Majorana $\pi$ modes to emerge at one end of a periodically driven topological superconductor. This finding was refined in~\cite{Zhou2018} by uncovering a mechanism to systematically generate an arbitrarily large number of zero and $\pi$ modes, albeit in a non-Majorana setting. In~\cite{Bomantara2020}, a similar mechanism for generating any number of Majorana zero and $\pi$ modes is unveiled and exploited to implement a quantum error correction scheme.

{{\it $\pi$ modes and Floquet time crystals.--} Floquet time crystals are recently discovered phases of matter that are uniquely found in periodically driven systems, first proposed independently by \cite{Sacha2015,Else2016}. In simple terms, FTCs are characterized by the existence of an observable $O$ that evolves periodically at an integer $>1$ multiple of the driving period under a generic physical initial state. Such a periodicity is further required to be robust under considerable perturbations and persist infinitely long in the thermodynamic limit. The proposal by \cite{Else2016} quickly garnered several experimental realizations~\cite{Zhang2017,Choi2017} in the following year. }

{A possible connection between Majorana $\pi$ modes and FTCs was first suggested in a work by Khemani \emph{et al.}~\cite{Khemani2016}. There, the phase diagram of a periodically driven Kitaev chain was obtained. By further highlighting the equivalence between such a system with a periodically driven Ising chain under the Jordan-Wigner transformation, it is found that a regime in which the former system supports Majorana $\pi$ modes corresponds to a regime in which the latter system belongs to the so-called $\pi$-spin glass phase. Such a phase is characterized by the clustering of all the Floquet eigenstates into pairs with $\pi/T$ quasienergy separation, which is responsible for the formation of a $2T$-periodic FTC. Indeed, a state satisfying $|\psi\rangle = a|\varepsilon\rangle +b|\varepsilon+\pi/T\rangle$, where $a,b\in \mathbb{C}$, returns to itself (up to a global phase factor) only after two periods, since $U|\psi\rangle = e^{-i \varepsilon T}\left(a|\varepsilon\rangle -b|\varepsilon+\pi/T\rangle\right)$. In the $\pi$-spin glass phase, a generic physical (product) state can be approximately written as a superposition of two Floquet eigenstates with quasienergy difference of $\pi/T$, and its period-doubling behavior could be observed by measuring the system's average magnetization~\cite{Else2016}.}

{In 2018, Bomantara and Gong \cite{Raditya2018} found that properties of an FTC can be obtained in a periodically driven Kitaev chain directly when it supports Majorana zero modes and Majorana $\pi$ modes simultaneously. Intuitively, the superposition of a Majorana zero mode and a Majorana $\pi$ mode yields an operator which exhibits a $2T$ periodicity (due to the $\pi$ phase difference between the two Majorana modes after one period). Moreover, due to the topological nature of Majorana modes, such an operator is naturally robust against errors and persists indefinitely in the limit of infinite chain. However, since the obtained $2T$-periodic operator resides at a system's boundary by construction, the resulting phase is often distinguished from the FTCs studied in~\cite{Else2016,Khemani2016} and is instead referred to as a ``boundary FTC"~\cite{Khemani2019}.}

{Remarkably, the Majorana-nature of zero modes and $\pi$ modes is unnecessary for the formation of boundary FTCs. Indeed, given a Floquet topological system which supports coexisting zero modes and $\pi$ modes, a topologically protected $2T$-periodic boundary state can be obtained from a superposition of a zero mode and a $\pi$ mode, which can further be captured by a local (boundary) observable. Such boundary FTCs have in turn been experimentally realized in various platforms, including acoustic waveguides \cite{zhu2022}, photonic waveguides \cite{Wang2022}, and superconducting qubits \cite{Zhang2022}.}

{{\it Beyond topological $\pi$ modes.--} The existence of $\pi$ modes and their physical implications elucidated above motivated the search for other unique properties of periodically driven systems that have no static counterpart. In particular, given that $\pi$ modes represent Floquet eigenstates or quasienergy excitations at half the driving frequency, the so-called $2\pi/k$ modes emerge as their natural generalizations. The latter were first introduced in a periodically driven chain of $Z_k$ parafermions~\cite{Sreejith2016} as its parafermionic quasienergy edge excitations at $1/k$ th the driving frequency. There, $Z_k$ parafermions $\psi_1$ and $\psi_2$ are a generalization of Majorana fermions that satisfy the algebra $\psi_1\psi_2 = e^{i 2\pi/k} \psi_2\psi_1$ and $\psi_1^k=\psi_2^k=1$.}

{In static systems, parafermion zero modes have been identified as superior candidates for topological qubits as they support a richer set of topologically protected quantum gates as compared with their Majorana counterparts~\cite{Hutter2016}. It can thus be envisioned that parafermion $2\pi/k$ modes may open a new approach in topological quantum computing, in the spirit of what has been achieved with Majorana $\pi$ modes. To the best of our knowledge, however, such an opportunity has remained unexplored at the time of writing this Perspective. The elusivity of parafermions and the difficulty in experimentally realizing them contribute to the scarcity of studies along this direction.}

{Recently, Bomantara~\cite{Bomantara2021} demonstrates the possibility of achieving a subset of parafermion $2\pi/k$ modes in a system of multiple interacting periodically driven Majorana chains. This is achieved by exploiting two important insights. First, the well-understood correlation between Majorana $\pi$ modes and period-doubling FTCs can be naturally generalized to establish a correspondence between parafermion $2\pi/k$ modes and period-$k$-tupling FTCs. Second, the structural similarity between the FTC model of~\cite{Else2016} and a quantum error correction model can be exploited to devise a family of period-$2^n$-tupling FTCs \cite{Raditya2021}. Therefore, by constructing a generalized Jordan-Wigner transformation that maps the quasi-1D spin-based FTC systems of \cite{Raditya2021} to the corresponding systems of Majorana fermions, the latter are found to host parafermion $2\pi/2^n$ modes at their boundaries ~\cite{Bomantara2021}. Moreover, the resulting systems only require the already attainable ingredients such as $p$-wave superconductivity, magnetic field, and Hubbard interaction.}

{Finally, a recent study~\cite{Bomantara2022} demonstrates that (non-parafermion) $\pi/2$ modes may also be obtained in the non-interacting setting. Specifically, they arise at the boundaries of a square-root Floquet topological system, i.e., a system whose corresponding Floquet operator squares to that of an existing Floqet topological system~\cite{Bomantara2022}. By repeating the square-root procedure of~\cite{Bomantara2022} multiple times, a system hosting $2\pi/2^n$ modes is also obtained. In \cite{Zhou2022}, such a square-root procedure is generalized to obtain any $k$th-root of a Floquet topological system, resulting in the emergence of any $2\pi/k$ modes. Around the same time, $\pi/2$ modes are successfully observed experimentally with acoustic waveguides \cite{Cheng2022}.}

{{\it Conclusion.--} In this Perspective, we have presented an overview of the discovery of topological $\pi$ modes as well as their physical significance in quantum computing and the understanding of an exotic phase of matter, i.e., the FTC. The recent proposals of $2\pi/k$ modes as the generalizations of $\pi$ modes have further been elucidated. In particular, with their physics slowly uncovered in recent studies and their experimental realizations becoming within reach of current technologies in acoustic\cite{zhu2022,Cheng2022} and superconducting circuits\cite{Zhang2022}, we expect that more ambitious studies exploring the applications of the exotic $2\pi/k$ modes shall appear in the near future. At the same time, the search for other novel features of periodically driven topological systems, especially in the presence of interaction effect, is expected to form a sustainable research direction for years to come. }

\bigskip

{\bf Conflict of interest:}
The authors declare that they have no conflict of interest.

\vspace{0.3cm}
{\bf Acknowledgements:}
%J.G. acknowledges fund support by the Singapore Ministry of Education Academic
%Research Fund Tier-3 Grant No. MOE2017-T3-1-001
%(WBS. No. R-144-000-425-592) and by the Singapore National Research Foundation
%Grant No. NRF-NRFI2017- 04 (WBS No. R-144-000-378-
%281).
J.G. acknowledges support from the Singapore National Research Foundation (grant award no. NRF2021-QEP2-02-P09).

\vspace{0.3cm}
{\bf Author contributions:}
All authors contributed to the planning and writing of this manuscript upon receiving the invitation from the journal. W.Z wrote a preliminary first draft. R.W.B extensively revised the draft.  J.G and R.W.B. finalized this manuscript.

\bibliography{reference}

%\begin{thebibliography}{68}%
%\bibitem{hasan2010colloquium}
%Hasan MZ, Kane CL. Colloquium: topological insulators. Rev Mod Phys 2010;82:3045.
%\end{thebibliography}%

\end{document}